%% file: paper.tex
\documentclass{acm_proc_article-sp}

\usepackage{xspace}
\usepackage{multirow}
\usepackage{url}
\usepackage{tabularx}

\newcommand{\R}{\mathbb{R}}
\newcommand{\xeon}{{\sc Xeon}\xspace}
\newcommand{\afat}{{\sc A15}\xspace}
\newcommand{\athin}{{\sc A7}\xspace}
\newcommand{\aft}{{\sc A15+nA7}\xspace}
\newcommand{\gemv}{{\sc gemv}\xspace}
\begin{document}

%\title{Iso-Performance, Iso-Power and Reliability\\ of Low Power big.LITTLE ARM Clusters\\ using Memory-Bounded Numerical Algorithms}
\title{Evaluating Asymmetric Multicore Systems-on-Chip\\ using Iso-Metrics}
%\subtitle{
%\titlenote{A full version of this paper is available as
%\textit{Author's Guide to Preparing ACM SIG Proceedings Using
%\LaTeX$2_\epsilon$\ and BibTeX} at

\numberofauthors{3} %  in this sample file, there are a *total*
\author{
% 1st. author
\alignauthor
Charalampos Chalios
       \affaddr{School of EEECS}\\
       \affaddr{Queen's University of Belfast}\\
       \affaddr{United Kingdom}\\
       \email{cchalios01@qub.ac.uk}
% 2nd. author
\alignauthor
Dimitrios S. Nikolopoulos
       \affaddr{School of EEECS}\\
       \affaddr{Queen's University of Belfast}\\
       \affaddr{United Kingdom}\\
       \email{d.nikolopoulos@qub.ac.uk}
% 3rd. author
\alignauthor
Enrique S. Quintana-Ort\'{\i}\\
       \affaddr{Depto. Ing. y Ciencia Comp.}\\
       \affaddr{Universitat Jaume~I, Castell\'on, Spain}\\
       \email{quintana@uji.es}
}
% There's nothing stopping you putting the seventh, eighth, etc.
% author on the opening page (as the 'third row') but we ask,
% for aesthetic reasons that you place these 'additional authors'
% in the \additional authors block, viz.
% Just remember to make sure that the TOTAL number of authors
% is the number that will appear on the first page PLUS the
% number that will appear in the \additionalauthors section.

\maketitle
\begin{abstract}
\input{abstract}
\end{abstract}

%A category including the fourth, optional field follows...

%\keywords{X,Y,Z} % NOT required for Proceedings

\input{body}

\section*{Acknowledgments}

E.S. Quintana-Ort\'{\i}
was supported by project TIN2011-23283 of
the MINECO and FEDER,
and the EU project FP7 318793 ``EXA2GREEN''.
This work was partially done while this author was
visiting 
Queen's University of Belfast.
We thank F.D. Igual, from {\em Universidad Complutense de Madrid}, for
his help with the Odroid board.

This research has been supported in part by the European Commission
under grant agreements FP7-323872 (SCoRPiO), FP6-610509 (NanoStreams)
and by the UK Engineering and Physical Sciences Research Council under
grant agreements EP/L000055/1 (ALEA), EP/L004232/1 (ENPOWER) and
EP/K017594/1 (GEMSCLAIM)

\bibliographystyle{plain}
%\bibliography{enrique,energy,asymmetric}
\bibliography{enrique,energy}

\end{document}

%% file: abstract.tex
The end of Dennard scaling has pushed power consumption into a first order concern
for current systems, on par with performance. 
%in the design of current systems, from commodity portable devices to HPC systems. In order to squeeze more performance
%out of the transistors, that keep following Moore's law, architects are trying to fit
%an increasing number of cores within the same socket. At the same time, the HPC world
%turned to high-end multiprocessor nodes in order to take advantage of the available parallelism,
%using sophisticated micro-architectures. This is no more a viable solution, as the power budget of
%systems' design is a limiting factor. 
As a result, near-threshold voltage computing (NTVC) 
has been proposed as a potential means to tackle the limited
cooling capacity of CMOS technology. 
Hardware operating in NTV consumes significantly less power,
at the cost of lower frequency, and thus reduced performance, as well as increased error rates. 
%At the same time, designers at all stacks need
%to address the problems that arise because of unreliability. 
In this paper, we investigate if a low-power systems-on-chip, consisting
of ARM's asymmetric big.LITTLE technology, can be
an alternative to conventional high performance multicore processors in terms of 
power/energy in an unreliable scenario. 
For our study, we use the Conjugate Gradient solver, an algorithm 
representative of the computations performed by a large range of scientific 
and engineering codes. 
%We use real platforms from both ends
%of performance to power-efficiency spectrum, in order to evaluate our methods. 

%% file: body.tex
\input{s1-introduction}

\input{s2-setup}

\input{s3-comparison}
\input{s3.1-tradeoff}
\input{s3.2-iso}

\input{s4-reliability}

\input{s5-remarks}

%% file: s1-introduction.tex
\section{Introduction}

The performance of today's computing systems is limited by 
the end of Dennard scaling~\cite{Den74} and
the cooling capacity of CMOS technology~\cite{%Dur13,Lec13,
Luc14}.
In response, CPU architectures turned towards multicore designs 
already in the middle of past decade, 
and power-saving techniques and mechanisms originally conceived for embedded 
and mobile appliances are being increasingly adopted by
desktop and server processors.
Near-threshold voltage computing (NTVC) is 
a promising power-saving technology 
to tackle the power wall by diminishing voltage (and slightly frequency)
of the processor at the cost of reducing hardware reliability~\cite{Kar13}.
The hope in NTVC is that the (close to) linear drop that is expected in performance from
the decay of frequency is compensated by cramming more cores into the same power
budget.
In addition, the increase in hardware concurrency can be exploited to integrate some sort
of algorithmic-based fault tolerance (ABFT) that addresses eventual data corruption
caused by operating with unreliable hardware.
 
In this paper, we investigate the performance, power and energy balance of
two representative low power ARM processors %, the Cortex-A7 and Cortex-A15, 
of a big.LITTLE system-on-chip (SoC), 
when applied to a memory-intensive numerical problem. 
Concretely, our analysis experimentally evaluates the
{\em iso-performance} and {\em iso-power} of quad-core ARM Cortex-A15 and Cortex-A7 {\em clusters}
against a conventional high performance Intel Xeon E5-2650 CPU, using
the Conjugate Gradient (CG) method~\cite{saad}. This memory-bounded algorithm 
for the solution of linear systems is particularly interesting as 
it is representative of the type of operations and performance attained by many other
scientific and engineering codes running in high performance computing 
facilities~\cite{AsaBCGHKPPSWY06}. %,hpcg}. 
As an additional contribution, we shed some light into the energy-saving potential of NTVC
under a realistic scenario.
For this purpose, we leverage a fault-tolerant variant of CG,
enhanced with a self-stabilizing (SS) recovery mechanism~\cite{Sao13},
to assess the practical energy trade-off 
between hardware concurrency, CPU frequency, and hardware error rate,
using the ARM big.LITTLE architecture as a case study. 

As part of related work, 
iso-energy-efficiency models are built in~\cite{6012831} 
in order to predict and balance energy and performance in large power-aware clusters, 
taking into account software characteristics. Compared to this, we focus 
on the trade-off between performance, power and energy for 
high-end multicore processors vs low power SoCs, designed mainly for 
embedded and mobile systems. Our goal is to answer whether it is possible to build
systems out of such power-efficient architectures that can match the performance of current 
throughput-oriented machines.
Similarly to us, the authors of~\cite{Göddeke2013132} study the use of power-efficient 
architectures in scientific applications. In this line, we take one step further, 
to make projections about the energy-efficiency of unreliable NTVC platforms and the use
of fault tolerance techniques to tackle the unreliability issues.

The rest of the paper is structured as follows.
In Section~\ref{sec:setup} we describe the experimental setup.
In Section~\ref{sec:comparison} we compare high performance vs low power architectures
using two different iso-metrics, and
in Section~\ref{sec:reliability} we determine the effect of unreliable hardware on 
the CG method.
Finally, we close the paper with a few remarks in Section~\ref{sec:remarks}.

%% file: s2-setup.tex
\section{Experimental Setup}
\label{sec:setup}

\input{table_hardware}

\subsection{The CG method}

The CG method is a key algorithm for the numerical solution of symmetric positive definite (SPD)
sparse and dense linear systems~\cite{saad} %,bekascurioni:2010} 
of the form
%Given a linear system 
$Ax=b$, where
$A \in \R^{n \times n}$ is SPD, $b \in \R^n$ contains the independent terms, 
and $x^n \in \R^n$ is the solution. 
%the CG method is algorithmically formulated 
%in Figure~\ref{fig:cg}. There, the user-defined parameters {\em maxres} and {\em maxiter}
%set upper bounds, respectively, on the relative residual for the computed approximation to 
%the solution $x_k$, and the maximum number of iterations.
%
%\begin{figure}
%\begin{center}
%\begin{tabular}{|l|}
%\hline
%\\ $x_0  :=  0$ // or any other initial guess
%\\ $r_0  :=  b-Ax_0$,~~~~
   %$d_0  :=  r_0$\\%,~~~~
   %$\beta_0 := r_0^Tr_0$,~~~~~~~~
   %$\tau_{0}:=\parallel r_{0}\parallel_2 = \sqrt{\beta_{0}}$,~~~~
   %$k := 0$
%\\ {\bf while} $(k < \mbox{\em maxiter})$ $\&$ $(\tau_{k} > %$& \hspace*{-2.5ex} $
%\mbox{\em maxres})$
%\\ ~~~~ $z_k:=Ad_k$ %& (SpMV)
%\\ ~~~~ $\rho_k:=\beta_k/{d_k^Tz_k}$  %& (dot)
%\\ ~~~~ $x_{k+1}:=x_k+\rho_k d_k$, %& (saxpy)	
%%\\ ~~~~ 
        %~~~~$r_{k+1}:=r_k-\rho_k z_k$ %& (saxpy)	
%\\ ~~~~ $\beta_{k+1}:= r_{k+1}^T r_{k+1}$ %& (dot)
%\\ ~~~~ $\alpha_{k}:=\beta_{k+1}/\beta_{k}$
%\\ ~~~~ $d_{k+1}:= r_{k+1} + \alpha_{k} d_k$ %& (scal+saxpy)
%\\ ~~~~ $\tau_{k+1}:=\parallel r_{k+1}\parallel_2 = \sqrt{\beta_{k+1}}$,
%%\\ ~~~~ 
        %~~~~$k := k+1$
%\\ {\bf end} \\
%\hline
%\end{tabular}
%\end{center}
%\caption{Algorithm for the CG method.}
%\label{fig:cg}
%\end{figure}
%
The cost of this iterative method is dominated by the matrix-vector multiplication 
(\gemv) with $A$ that is computed per iteration.
%$z_k:=Ad_k$.  
For a matrix $A$ with $n_z$ nonzero entries,
this operation roughly requires 
$2n_z$ floating-point arithmetic operations (flops).
Additionally, each iteration involves a few vector operations
% (for the updates of $x_{k+1}$, $r_{k+1}$, $d_{k+1}$, and the computation 
%of $\rho_{k}$ and $\beta_{k+1}$) 
that cost $O(n)$ flops each.

For our evaluation, we employ {\sc ieee} 754 real double-precision arithmetic 
and stop the iteration when the relative residual of the approximated solution is below 1.0e$-$8.
Furthermore, we consider only problems with dense $A$ and,
for simplicity, we do not exploit the symmetric structure of the matrix.
Under these conditions, we estimate the cost per iteration of CG 
to be $2n^2$ flops (i.e., we neglect the lower cost of
the vector operations). Moreover, for efficiency, we leverage
multi-threaded implementations of the \gemv kernel in 
Intel MKL (version 11) for the Intel-based CPU, and ATLAS (version 3.8.4)
for the ARM-based cores.

\subsection{Target architectures and scenarios}

The experiments in this paper were performed using three different CPUs.
The first one, hereafter \xeon, 
is a high-performance but power-hungry Intel Xeon E5-2650 socket with
16~GBytes of DDR3-1333~MHz RAM.
The alternative low-power architectures, \afat and \athin, are two ARM quad-core clusters
embedded into an Exynos5 system-on-chip (SoC) of an ODROID-XU board, sharing 2~Gbytes
of DDR3-800~MHz RAM. 
Table~\ref{tab:hardware} offers the most important features of these CPU architectures.
There, the ``Stream bandwidth'' column reports the
memory bandwidth measured using the {\tt triad} test of the {\tt stream} 
benchmark\footnote{\url{http://www.cs.virginia.edu/stream}} on the highest number of
cores available in the sockets. The ``Roofline GFLOPS'' column
corresponds to the theoretical upper bound on the computational performance
(in terms of GFLOPS, or billions of flops per second) dictated by the 
{\em roofline model}. %~\cite{Williams:2009:RIV:1498765.1498785}.

For the evaluation, we investigate
different scenarios that vary in the number of cores (from 1 up to the maximum), 
the CPU frequency, and the problem size. For simplicity, we only consider 
two CPU frequencies (lowest and highest, in particular discarding  Intel's turbo-mode)
for each architecture; and two problem dimensions:
an ``on-chip'' case that occupies much of the last level of cache (LLC), $n$=1,024 on \xeon, 
$n$=512 on \afat and $n$=256 on \athin; and an ``off-chip'' counterpart that clearly
exceeds the capacity of the LLC, with 
$n$=4,096 on \xeon, 
$n$=1,024 on \afat and
$n$=512 on \athin.

%% file: table_hardware.tex
\begin{table*}[tbh!]
\begin{center}
{\scriptsize
\begin{tabular}{|l|l||c|c|c|c|c|c|c|}
\hline
&                  & \#Cores & Frequency   &  LLC:          & TDP & Peak  mem.& Stream mem.& Roofline
\\
       &                   &   & range     & level, type,   & (W) & bandwidth & bandwidth  & GFLOPS
\\
Acron. & CPU socket/cluster&   & (GHz)     & size (Mbytes)  &     & (GBytes/s)& (Gbytes/s) &       
\\ \hline \hline
\xeon & Intel Xeon E5-2650 & 8 & 1.2--2.0  &  L3, shared, 20& 95  & 51.2      & 44       & 11 
\\ \hline
\afat & ARM Cortex-A15     & 4 & 0.8--1.6  &  L2, shared, 2~~& N/A & N/A       & 5.4     & 1.35
\\ \hline
\athin& ARM Cortex-A7      & 4 & 0.5--1.2  &  L2, shared, 0.5~~& N/A & N/A       & 2.07      & 0.51
\\ \hline
\end{tabular}
}
\caption{Hardware specifications of the target architectures.}
\label{tab:hardware}
\end{center}
\end{table*}

%% file: s3-comparison.tex
\section{High Performance vs Low Power}
\label{sec:comparison}

In this section, we perform an experimental evaluation
of the target CPU architectures, using 
the CG method (implemented on top of optimized multi-threaded versions of MKL and ATLAS), from the points of view
of performance, power dissipation, and energy consumption.
The purpose of this analysis is to expose the trade-offs between these
three metrics, for a memory-bound method such as CG, on these particular
architectures, with the ultimate goal of answering two key questions:
\vspace*{-3ex}
\begin{itemize}
\itemsep=0pt\parskip=0pt
\item
{\sc Q1} ({\em Iso-performance}): Can we attain the performance of the Intel Xeon
CPU with the low power ARM clusters while yielding a more power-efficient solution?
\item
{\sc Q2} ({\em Iso-power}): What is the performance that can be attained using
the low power ARM clusters within the power budget dictated by the
Intel Xeon socket?
\end{itemize}

%% file: s3.1-tradeoff.tex
\subsection{Trade-offs}

\begin{figure*}[tbh!]
\begin{center}
\includegraphics[height=4.3cm,width=0.8\textwidth]{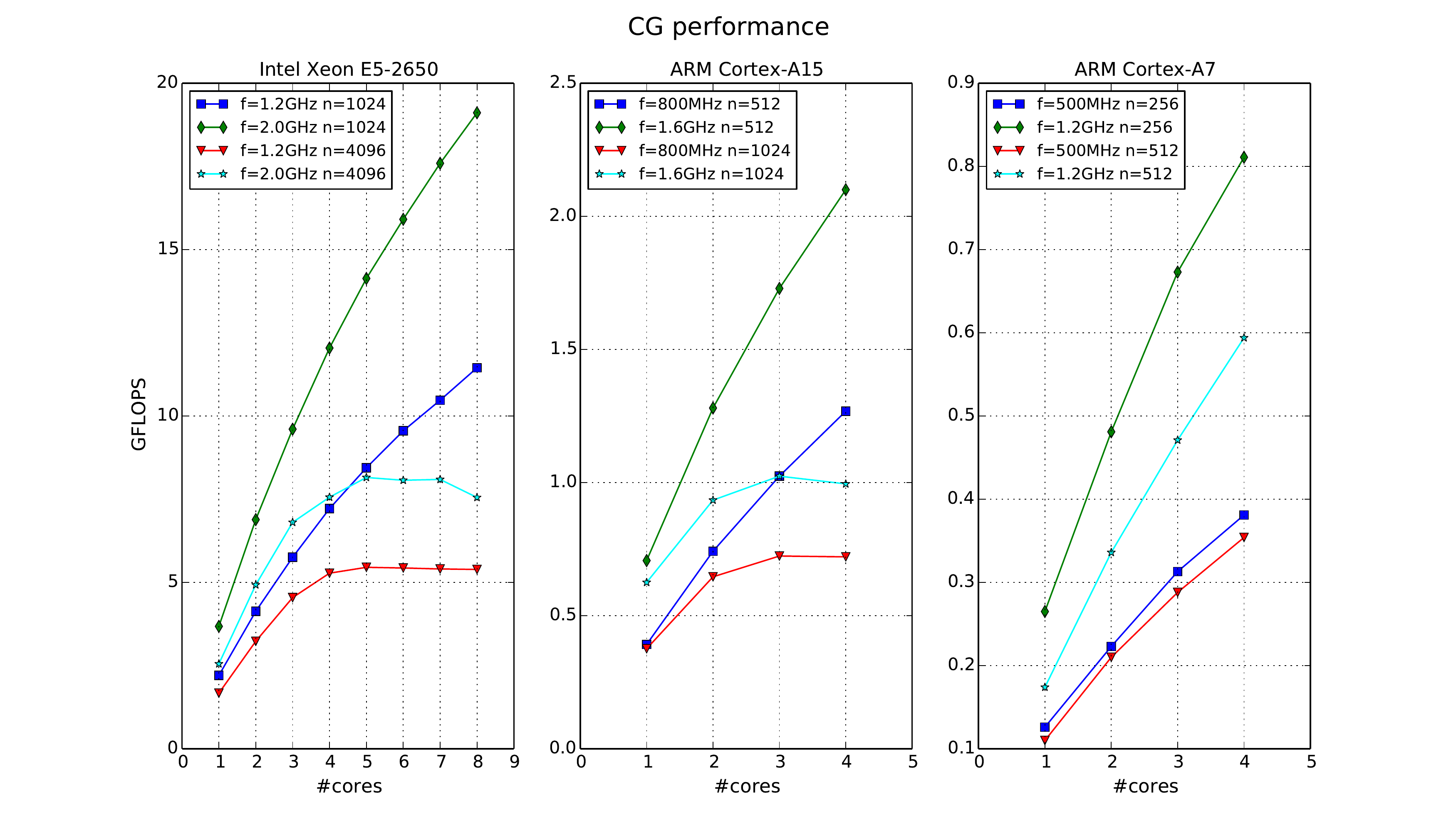}\\
\includegraphics[height=4.3cm,width=0.8\textwidth]{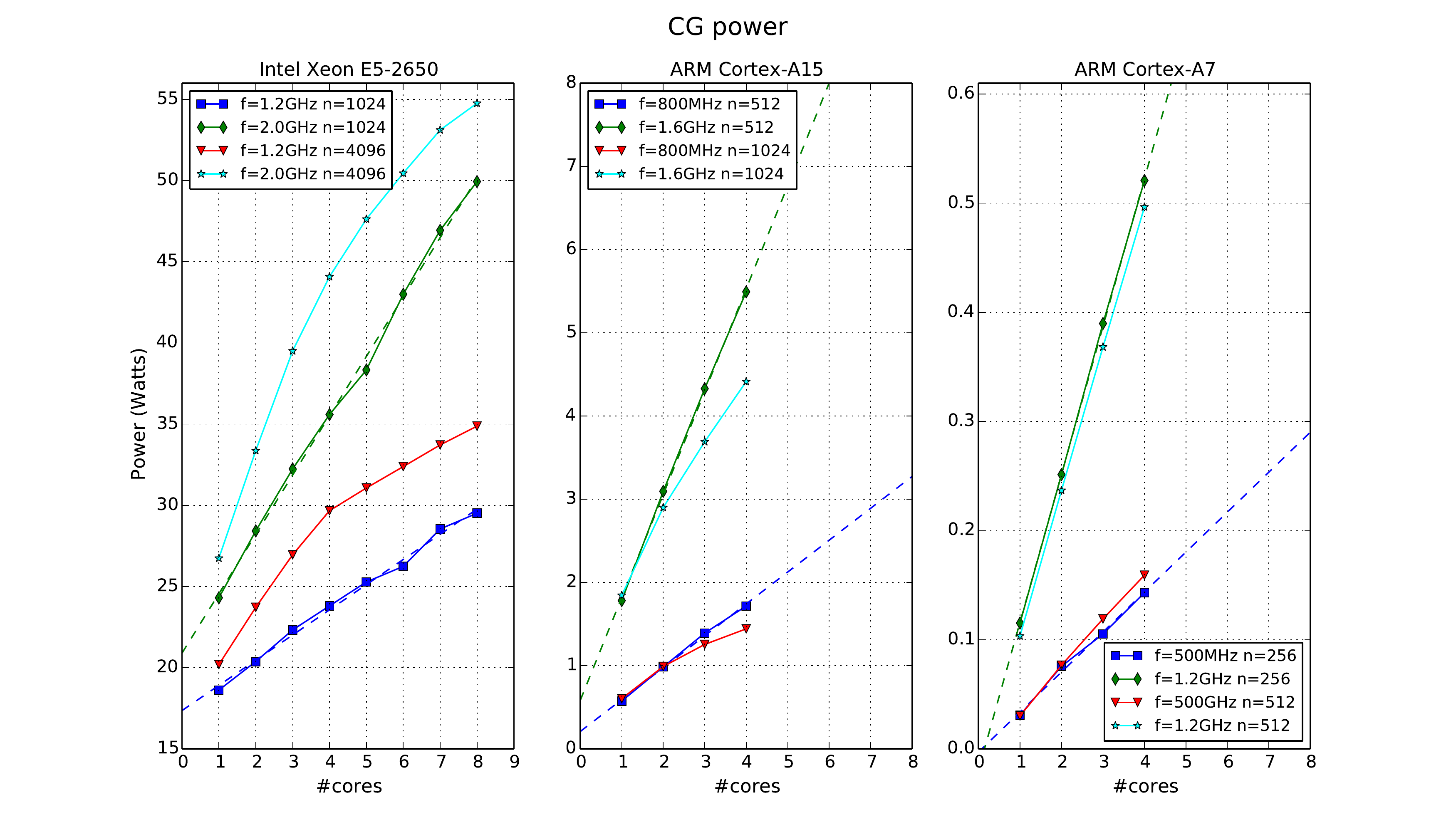}\\
\includegraphics[height=4.3cm,width=0.8\textwidth]{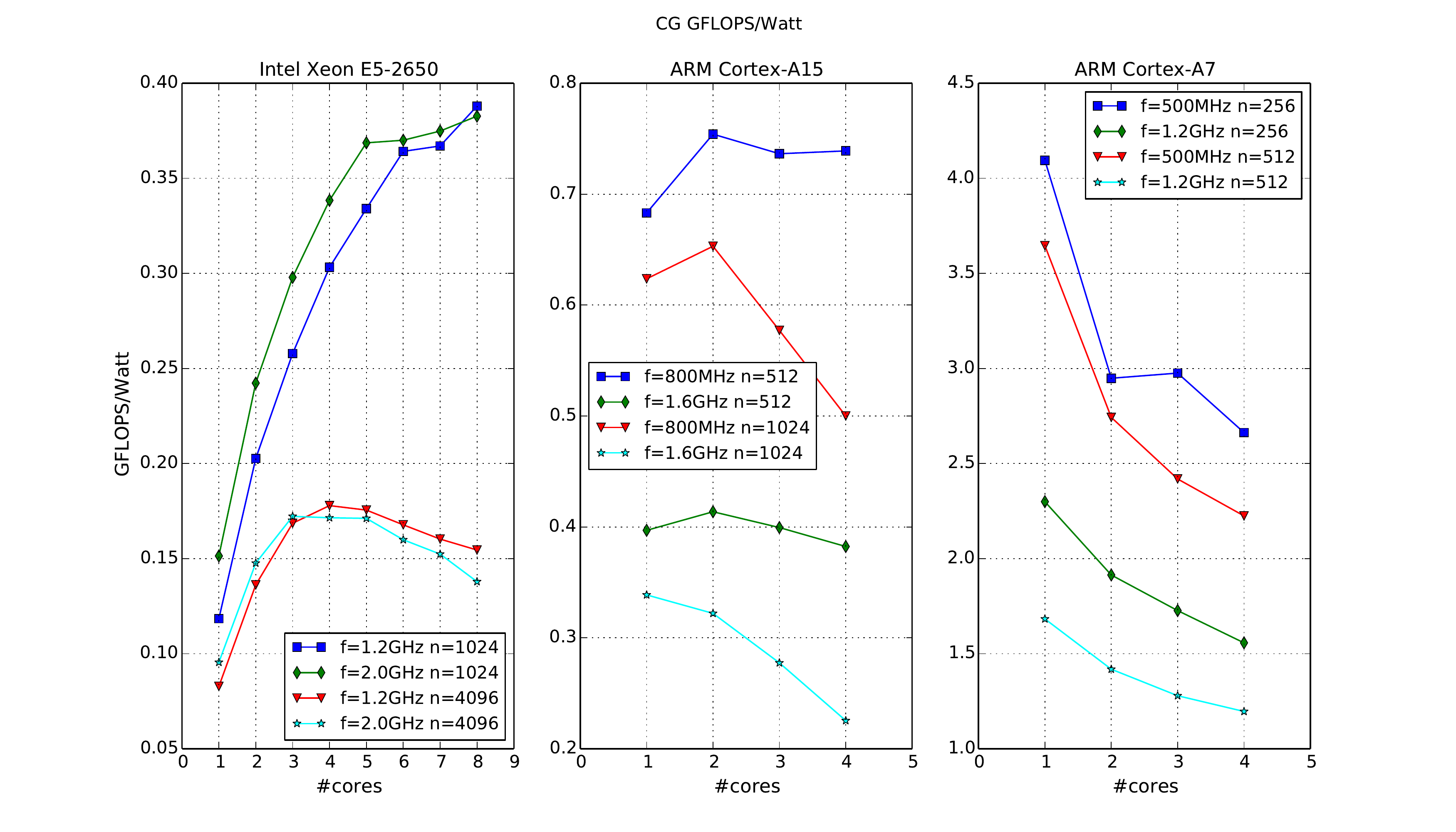}
\caption{Evaluation of performance, power and energy on the target architectures
         using multi-threaded implementations of the CG method on both the on-chip and off-chip problems.}
\label{fig:comparison}
\end{center}
\end{figure*}

\input{table_comparison}

Figure~\ref{fig:comparison} reports the results from the 
evaluation of the multi-threaded CG implementations, 
from the points of view of performance (in GFLOPS), power dissipation (W) and 
energy efficiency (GFLOPS/W), using both
on-chip and off-chip problems. 
We note that an evaluation in terms of GFLOPS and GFLOPS/W allows a comparison 
of these metrics for problems of varying size, 
which require a different number of flops. % to be solved, in part
%because they may also require a distinct number of iterations to converge.

We start by distinguishing between the two scenarios corresponding to on-chip and off-chip problems.
For brevity, we will focus hereafter in the former case, noting that, in the latter, the performance
on \xeon and \afat is clearly limited by the memory bandwidth, offering considerably lower figures
on all three metrics.
The same memory bottleneck is not visible for \athin though, likely because the
multi-threaded implementation of the matrix-vector multiplication in ATLAS does not extract all the  performance
of this architecture.

Table~\ref{tab:comparison} offers numerical results for the on-chip problems. Our comments 
to these results are organized in three axis: \#cores, frequency and
architecture (configuration parameters) as well as three perspectives (metrics).
Let us commence by putting the light on the \#cores.
From the point of view of concurrency, increasing \#cores produces fair speed-ups,
which interestingly are quite close for all three architectures
independently of their frequency; e.g., the use
of 4 cores on \xeon, \afat and \athin produces speed-ups between 2.8 and 3.4 for any of the
two frequencies.
From the perspective of power, a linear regression fit to the data shows 
a high value of the $y$-intercept for \xeon, which basically corresponds to
static power, and can be explained by its large LLC, the complex pipeline, the large
area dedicated to branch prediction, etc.
Compared with this, \afat and \athin exhibit much lower static power, reflecting the simpler
design of this CPU clusters.
This difference between the Intel- and ARM-based  architectures 
has a major impact on the energy where, e.g.,
increasing the \#cores on \xeon results in shorter execution time and, due to the large static
power, a visible positive effect on energy efficiency (GFLOPS/W). This is a clear indicator of
the potential benefits of a ``race-to-idle'' policy applied to this architecture. The effect of increasing \#cores on \afat and \athin 
is more imprecise, due
to the low fraction that the static power represents. % on these architectures. 

We continue next with the analysis of frequency. Independently of the number of cores,
the effect of this parameter on performance is perfectly linear for \xeon but sublinear
for \afat, where doubling the frequency only improves performance by a factor of about 1.7$\times$;
and slightly higher for \athin, where raising the frequency from 0.5 to 1.2~GHz (a factor of
2.4$\times$) results in an increase of performance 2.1$\times$.
The effect of frequency on power is sublinear for \xeon (a factor between 1.30--1.69$\times$,
depending on the number of cores)
and superlinear for both \afat (3.12--3.20$\times$) and \athin (3.66--3.71$\times$).
The net effect of the variations of time and power with the frequency is that, on \xeon,
increasing the frequency slightly improves energy efficiency (race-to-idle) while on
the ARM-based clusters it reduces it by a factor close to 50\% for \afat and 64\% for \athin.

Finally, we observe some general differences between the CPU architectures:
the power hungry 8-core Intel CPU produces significantly higher
performance rates (and, therefore, shorter execution times) than the ARM clusters,
at the expense of a much higher dissipation rate and lower energy efficiency.
The differences between \afat and \athin follow a similar pattern, with higher performance
in the former in exchange for higher power draft/lower energy efficiency.

%% file: table_comparison.tex
\begin{table*}[tbh!]
\begin{center}
%{\footnotesize
{\scriptsize
\begin{tabular}{|l|c||c|c|c|c|c|c|}
\hline
CPU   & Freq. &\#cores & Time per  & Performance & Speed-up & Power      & Energy     \\
      & (GHz) &        & iter. (ms)& (GFLOPS)    &          & (W)        & (GFLOPS/W) \\ \hline \hline
\multirow{16}{*}{\xeon} & 
\multirow{8}{*}{1.2}   
              & 1      & 0.89      & ~2.21  & 1.0      & 18.6       & 0.12     \\ %\cline{3-8}
      &       & 2      & 0.47      & ~4.12  & 1.9      & 20.4       & 0.20     \\ %\cline{3-8}
%      &       & 3      & 0.34      & ~5.75  & 2.6      & 22.3       & 0.26     \\ %\cline{3-8}
      &       & 4      & 0.27      & ~7.21  & 3.3      & 23.8       & 0.30     \\ %\cline{3-8}
%      &       & 5      & 0.23      & ~8.44  & 3.8      & 25.3       & 0.33     \\ %\cline{3-8}
      &       & 6      & 0.21      & ~9.55  & 4.3      & 26.2       & 0.36     \\ %\cline{3-8}
%      &       & 7      & 0.19      & 10.47  & 4.7      & 28.5       & 0.37     \\ %\cline{3-8}
      &       & 8      & 0.17      & 11.44  & 5.2      & 29.5       & 0.39     \\ \cline{2-8}
&
\multirow{8}{*}{2.0}   
              & 1      & 0.53      & ~3.67  & 1.0      & 24.3       & 0.15     \\ %\cline{3-8}
      &       & 2      & 0.28      & ~6.88  & 1.9      & 28.4       & 0.24     \\ %\cline{3-8}
%      &       & 3      & 0.20      & ~9.60  & 2.6      & 32.2       & 0.30     \\ %\cline{3-8}
      &       & 4      & 0.16      & 12.04  & 3.3      & 35.5       & 0.34     \\ %\cline{3-8}
%      &       & 5      & 0.13      & 14.13  & 3.8      & 38.3       & 0.37     \\ %\cline{3-8}
      &       & 6      & 0.12      & 15.91  & 4.3      & 42.9       & 0.37     \\ %\cline{3-8}
%      &       & 7      & 0.11      & 17.59  & 4.8      & 46.9       & 0.37     \\ %\cline{3-8}
      &       & 8      & 0.10      & 19.11  & 5.2      & 49.9       & 0.38     \\ \hline \hline
\multirow{8}{*}{\afat} & 
\multirow{5}{*}{0.8}   
              & 1      & 1.26      & ~0.39  & 1.0      & 0.57       & 0.68     \\ %\cline{3-8}
      &       & 2      & 0.66      & ~0.74  & 1.9      & 0.98       & 0.75     \\ %\cline{3-8}
%      &       & 3      & 0.48      & ~1.02  & 2.7      & 1.39       & 0.74     \\ %\cline{3-8}
      &       & 4      & 0.39      & ~1.26  & 3.2      & 1.71       & 0.74     \\ \cline{2-8}
& \multirow{5}{*}{1.6}   
              & 1      & 0.70      & ~0.70  & 1.0      & 1.78       & 0.40     \\ %\cline{3-8}
      &       & 2      & 0.40      & ~1.28  & 1.8      & 3.09       & 0.41     \\ %\cline{3-8}
%      &       & 3      & 0.29      & ~1.72  & 2.4      & 4.32       & 0.40     \\ %\cline{3-8}
      &       & 4      & 0.25      & ~2.10  & 2.8      & 5.49       & 0.38     \\ \hline \hline
\multirow{8}{*}{\athin} & 
\multirow{5}{*}{0.5}   
              & 1      & 0.98      & ~0.12  & 1.0      & 0.03       & 4.09     \\ %\cline{3-8}
      &       & 2      & 0.56      & ~0.22  & 1.8      & 0.07       & 2.95     \\ %\cline{3-8}
%      &       & 3      & 0.39      & ~0.31  & 2.5      & 0.10       & 2.98     \\ %\cline{3-8}
      &       & 4      & 0.32      & ~0.38  & 3.0      & 0.14       & 2.66     \\ \cline{2-8}
&
\multirow{5}{*}{1.2}   
              & 1      & 0.48      & ~0.26  & 1.0      & 0.11       & 2.30     \\ %\cline{3-8}
      &       & 2      & 0.26      & ~0.48  & 1.2      & 0.25       & 1.91     \\ %\cline{3-8}
%      &       & 3      & 0.19      & ~0.67  & 2.5      & 0.38       & 1.73     \\ %\cline{3-8}
      &       & 4      & 0.16      & ~0.81  & 2.9      & 0.52       & 1.56     \\ \hline
\end{tabular}
}
\caption{Evaluation of performance, power and energy on the target architectures
         using multi-threaded implementations of the CG method on the on-chip problems.}
\label{tab:comparison}
\end{center}
\end{table*}

%% file: s3.2-iso.tex
\subsection{Analysis of iso-metrics}

\begin{figure}
\begin{center}
\includegraphics[height=4.3cm,width=\columnwidth]{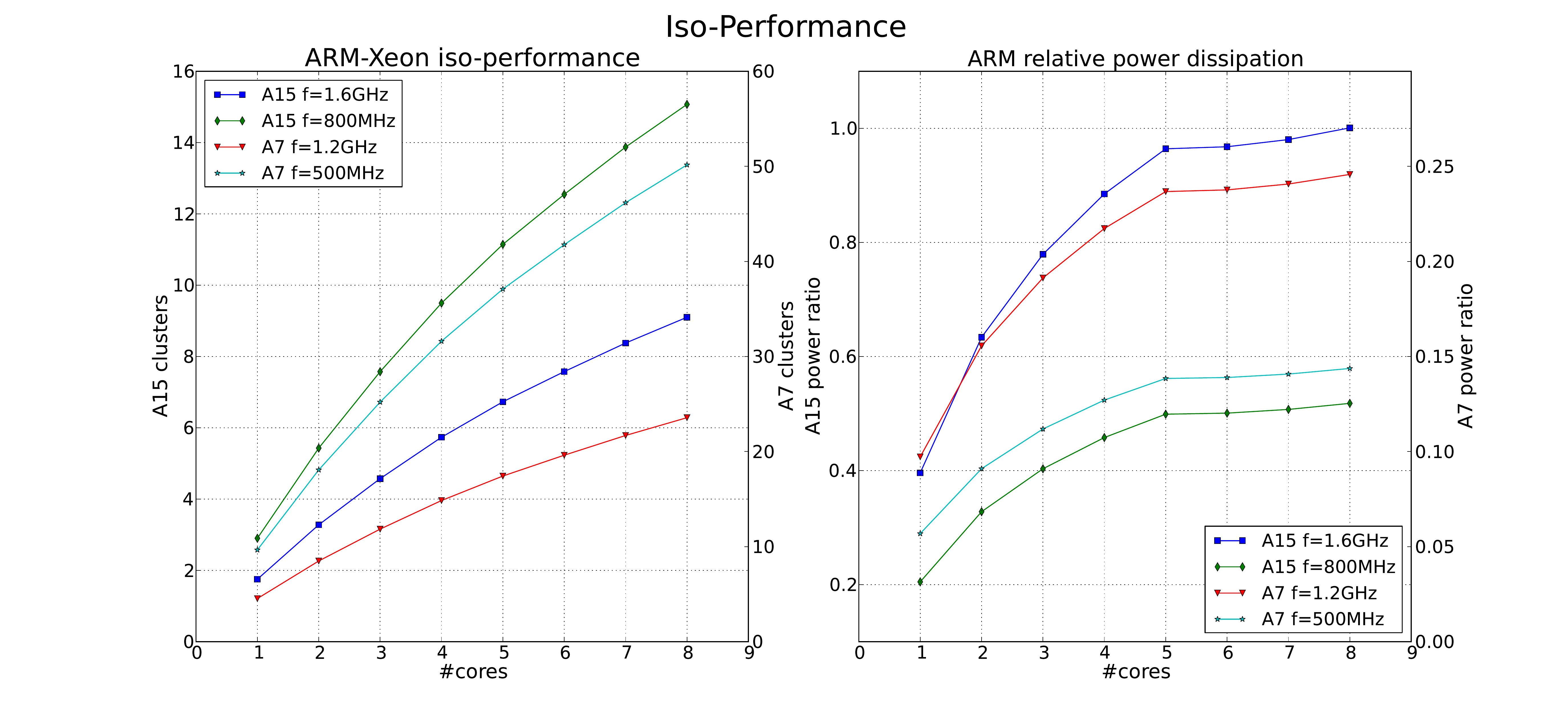}
\caption{Evaluation of iso-performance. Left: Number of \afat or \athin clusters o match the performance
         of a given number of \xeon cores at 2.0~GHz. Right: Comparison of power rates dissipated for configurations delivering
         the same performance.}
\label{fig:isoperformance}
\end{center}
\end{figure}

We open the following study by noting that %, given the bottleneck on the memory bandwidth for the operations underlying the
%CG method, 
the questions {\sc Q1} (iso-performance) and {\sc Q2} (iso-power) formulated at the beginning of this section can be analyzed in a different
number of configurations/scenarios. Here we select one that we find specially appealing.
Concretely, for {\sc Q1} we consider the performance of 1--8 cores from \xeon, at 2.0~GHz, as the objective,
and then we evaluate how many clusters (consisting of \afat or \athin 
and operating at either the lowest or the highest
frequencies) are necessary to match the reference performance. 
Question {\sc Q2} is the iso-power counterpart of the Q1, with
the power budget reference fixed by the power rate
of 1--8 cores from \xeon, at 2.0~GHz. Because of the scalability issue, in all cases we employ the 
performance and power rates observed when operating with on-chip problems.

The left-hand side plot in Figure~\ref{fig:isoperformance} reports the results from the iso-performance study, exposing that, in order to attain the performance of 8~cores from \xeon (2.0~GHz), it is necessary to use about 9.1~\afat clusters (i.e., quad-cores)
at 1.6~GHz or more than 50.2~\athin clusters at 0.5~GHz! (Note the different scales of the $y$-axis depending on the type
of cluster). Now, we recognize that in such comparison we implicitly introduce a simplifying assumption in favour of the
ARM CPUs. In particular, for the on-chip problem on \xeon, the dimension $n$=1,024. 
Now, in order to solve the same problem on a multi-socket ARM platform, data and operations 
have to be partitioned among and mapped to the clusters, %In consequence, depending on the ratio between the
%problem dimension and the number of clusters, 
incurring into overhead due to communication. 
For the CG method, we can expect that this additional cost
comes mostly from the reduction vector operations
(analogous to a synchronization). %On the other hand, the matrix-vector multiplication is highly parallel. 
Also, there is a certain overhead due to operating with a smaller problem size per core. 

The right-hand side plot in Figure~\ref{fig:isoperformance} illustrates the ratio between the power rates dissipated
by four configuration ``pairs'' that attain the same
performance, with one of the components of these pairs being \xeon and  
the other \afat or \athin, at either the lowest or the highest frequency.
Following with the previous examples, 8~cores from \xeon (at 2.0~GHz) deliver the same performance as  9.1~clusters from 
\afat at 1.6~GHz, and they draw basically the same power 
rate (a ratio of 1.001 between the two). On the other hand,
using 50.2~clusters of \athin at 0.5~GHz only requires a fraction of the power rate
dissipated by \xeon, concretely 14\%.

\begin{figure}
\begin{center}
\includegraphics[height=4.0cm,width=0.9\columnwidth]{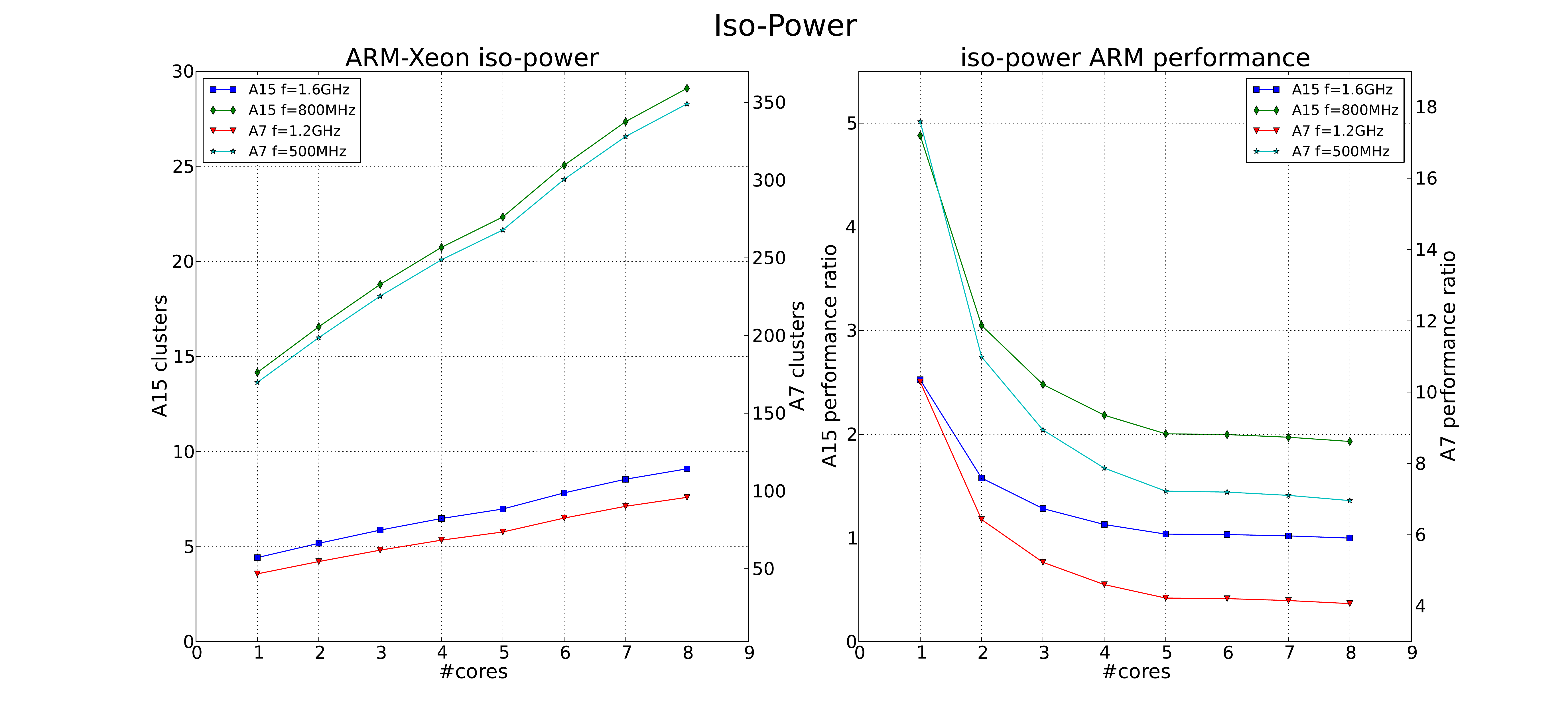}
\caption{Evaluation of iso-power. Left: Number of \afat or \athin clusters that match the power dissipated
         by a given number of \xeon cores at 2.0~GHz. Right: Comparison of performance rates attained for configurations dissipating
         the same power rate.}
\label{fig:isopower}
\end{center}
\end{figure}

Figure~\ref{fig:isopower} displays the results from the complementary study on iso-power.
The plot in the right illustrates that with the power budget of 1--8 \xeon cores, 
it is possible to accommodate a moderate number of \afat clusters or a very large volume
of \athin ones. The performance ratio between these ARM-based clusters with respect
to the \xeon, in the left plot, reveals decreasing gains with the number of \afat clusters and
a performance tie with respect to 4 or more \xeon cores. The ratio also decays for the 
\athin clusters, but in this case it is stabilized around a factor of 7.

Note that not all ARM-based configurations considered in the iso-performance and iso-power study have the same 
on-chip memory capacity (iso-capacity) as \xeon. In particular, given that the LLC 
for the latter is 20~MBytes, 
one need at least 10 \afat clusters and 40 \athin clusters to be in an iso-capacity scenario from the on-chip memory point of view.

We conclude this section by noting that a study of the energy efficiency ratio under the conditions imposed by
{\sc Q1} or {\sc Q2} does not contribute new information. For example, given that {\sc Q1}
basically relates the GFLOPS/W of two architectures with equal GFLOPS rates, an evaluation of energy efficiency
boils down to the analysis of the power ratio.

%% file: s4-reliability.tex
\section{Energy Cost of Reliability}
\label{sec:reliability}

The experiments and analysis in this section aim to expose the
potential impact on energy exerted by a technique that, like NTVC,
trades off lower CPU (voltage and) frequency and, therefore, more reduced power
consumption, for increased hardware concurrency and failure rate.
In order to perform this study in a realistic scenario, we raise the following
considerations:
\vspace*{-3ex}
\begin{itemize} 
\itemsep=0pt\parskip=0pt
\item We employ a tuned variant of our multi-threaded implementations of the CG method, 
      equipped with a SS recovery mechanism~\cite{Sao13} to cope with silent data corruption 
      introduced by unreliable hardware.
      Following the experiments in~\cite{Sao13}, the SS part is activated every~10 
      iterations of the CG method, and must be performed in reliable mode. 
      From the computational point of view, the major difference between an SS iteration and
      a ``normal'' CG one is that the former performs a total of two \gemv instead of only one.
      However, these two \gemv can be performed simultaneously, as they both involve $A$. Therefore,
      for a memory-bound operation like \gemv, we can consider that in practice, the 
      two types of iterations share the same computational cost.
\item To accommodate a reliable+unreliable execution, 
      we consider an  ``ideal'' 
      multi-socket big.LITTLE SoC consisting of
      a single quad-core \afat cluster plus  several \athin clusters.
      Here, \afat operates at the highest
      frequency, is considered to be reliable, and applies the SS mechanism.
      On the other hand, the \athin clusters operate at the lowest frequency, 
      represent the unreliable hardware, and are used to compute the
      normal CG iterations. We will refer to this SoC as \aft,
      and we will use data corresponding to on-chip problems for
      all the experimentation.
\item The convergence rate of the CG iteration depends on the condition number of matrix
      $A$~\cite{saad}. Under certain conditions, the convergence of the SS variant
      degrades logarithmically with the error rate~\cite{Sao13}.
      Silent data corruption is assumed to occur during \gemv, producing one or more
      bit flips into any of its results, and propagates 
      from there to the rest of the computations. The convergence rate of the SS variant also 
      depends mildly on whether the bit flips are bounded to the sign/mantissa or can
      affect also the exponent.
\end{itemize}
Under these conditions, 
we next perform an experimental analysis of the energy gains
that such a reliable.unreliable big.LITTLE SoC features, comparing it with a reliable
single quad-core \afat cluster operating at the highest frequency under iso-performance
and iso-power conditions. 

\begin{figure}
\begin{center}
\includegraphics[height=4.0cm,width=\columnwidth]{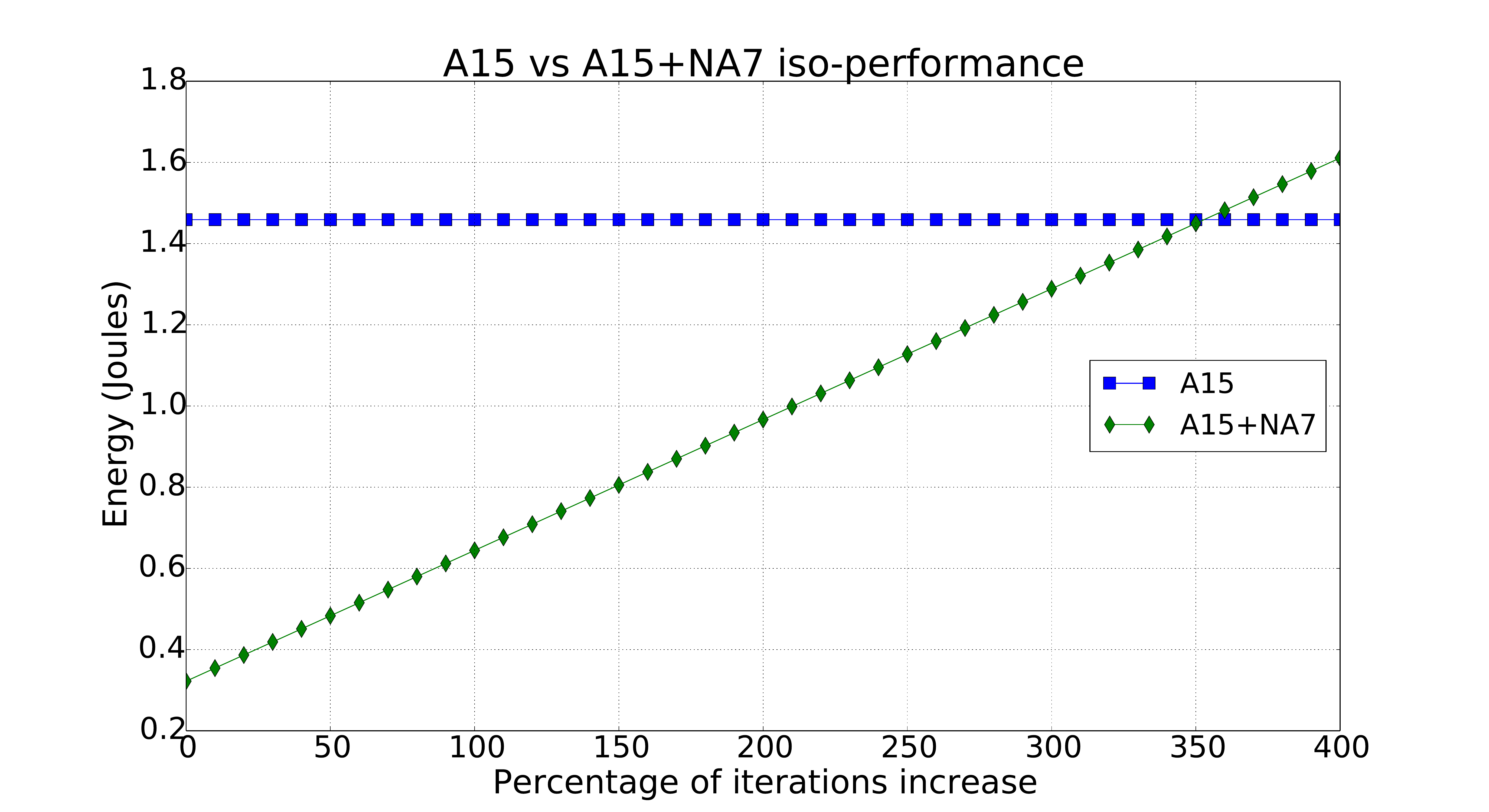}
%\vspace*{4.5cm}
\caption{Iso-performance ETS for the original CG method executed by \afat at the highest frequency
         (reliable mode) and the SS variant of CG executed by \aft under unreliable
         conditions which degrade convergence.}
\label{fig:isoperformance_rate}
\end{center}
\end{figure}

We commence with the iso-performance study. The first goal is to find how many
\athin clusters must be involved during the execution of the CG iterations so that,
when combined to build \aft
with a single \afat cluster for the execution of SS iterations (10\% of the
total), the performance that is obtained matches that of a single \afat cluster operating
at the highest frequency (i.e., 2.1 GFLOPS; see Table~\ref{tab:comparison}).
A little arithmetic gives an answer of 5.51~A7 clusters, which we will round to
6~A7 clusters, at the price of attaining a performance 
slightly above the reference objective (concretely, 2.28~GFLOPS). We can next compare
the power dissipation rate of the two cases: 5.49~W for \afat and 
1.31~W for \aft. Next, the GFLOPS rates for each two configurations, combined
with the cost per iteration ($2n^2$) and the number of iterations required for
convergence in the $n$=512 case, offers the execution times (slightly smaller for \aft,
because of the rounding). A combination of time with the previous power rates 
thus offers the {\em energy-to-solution} (ETS), i.e., how much energy (in Joules) 
is required to solve the same problem, on each architecture, in absence of errors
(though \aft applies the SS mechanism nonetheless). 
Finally, in Figure~\ref{fig:isoperformance_rate}, we compare
the ETS attained by original CG method, executed
in a reliable environment, against that of the SS variant, under unreliable conditions,
as the convergence degrades a certain percentage of iterations due to errors.
These results explicitly expose the energy gains that can be expected from operating
with simpler low power cores, at low frequencies, for this particular application,
with \aft outperforming \afat in terms of ETS when the degradation
incurs in up to 340\% more iterations. 

\begin{table}
\begin{center}
%{\footnotesize
{\scriptsize
\begin{tabular}{|l|c|c|c|}
\hline
                & \multicolumn{3}{c|}{\aft} \\ 
Case study      & \#\athin clusters & GFLOPS  & Power\\ \hline \hline
iso-performance & ~5.51 & ~2.09 & 1.24 \\
iso-power       & 38.85 & 13.49 & 5.44 \\
iso-capacity    & ~4    & ~1.57 & 1.05 \\
\hline
\end{tabular}
}
\caption{Comparison of \aft to \afat under iso-performance, iso-power and iso-capacity
conditions.}
\label{tab:unrel_iso}
\end{center}
\end{table}

We also perform an analogous study from the point of view of iso-power; that is, 
we set the power dissipated by the \afat cluster, at the highest frequency, as the reference (5.49~W; 
see Table~\ref{tab:comparison}),
and then we derive how many \athin clusters can be embedded into \aft within the same
power budget, with the answer being 38.85. 
This exercise will, eventually, produce the same ETS as the iso-performance 
analysis. This is to be expected, since any increase of \#\athin clusters in
\aft yields an proportional increase of its GFLOPS rate, or equivalently
an inversely proportional decrease in execution time. 
Simultaneously, the power dissipation will be increased in the same proportion,
yielding the same ETS.

To conclude this section, we focus on the iso-capacity problem. For this case-study,
we require the aggregated LLC of the \athin clusters in \aft to be equal that of \afat.
Figure~\ref{fig:comparison} shows that it is important that the data involved in the computation 
fit in the LLC so that the performance will scale with \#cores. 
Now, \afat includes a 2MB LLC cache, which can hold a 
problem size of $n$=512 for CG. Therefore, four \athin clusters
match the LLC capacity of a single \afat cluster
(see Table~\ref{tab:hardware}).  
In conclusion, we can build an \aft system which can solve the same problem size as
\afat, with a throughput of 1.57~GFLOPS, i.e. 1.33$\times$ slower than \afat, 
but dissipates 5.22$\times$ less power.
The iso-performance, iso-power and iso-capacity results are summarized in Table~\ref{tab:unrel_iso}.

%\include{table_unrel_iso}

%The answer in this case is XX~\athin clusters, 
%which we approximate as YY. Following a similar process as that just explained
%for the iso-performance, we arrive to the comparison between the ETS of these
%two configurations in Figure~\ref{fig:isopower_rate}.

%\begin{figure}
%\begin{center}
%\includegraphics[height=4.9cm,width=\columnwidth]{results/figs/iso-performance.pdf}
%\vspace*{4.5cm}
%\caption{Iso-power ETS for the original CG method executed by \afat at the highest frequency
%         (reliable mode) and the SS variant of CG executed by \aft under unreliable
%         conditions that degree convergence.}
%\label{fig:isopower_rate}
%\end{center}
%\end{figure}

%% file: s5-remarks.tex
\section{Conclusions and Future Work}
\label{sec:remarks}

The requirement for energy-efficient systems on the road towards 
Exascale systems asks for more power-efficient hardware designs. 
In this paper, we turn to the embedded and mobile world and investigate whether platforms 
from that domain can be used to build systems for HPC
applications with better energy-to-performance ratios. 
Concretely, we show that, in principle, it is possible to use 
power-efficient ARM clusters in order to match the performance of a high-end Intel Xeon processor 
while operating, in a worst-case scenario, at the same power budget. 
Conversely, it is also possible to use a rather large number of ARM clusters, fit 
into the power budget of one Intel Xeon processor, and attain higher performance.

As a second contribution, we experiment with a reliable CG execution in an \afat cluster 
versus an execution of a self-stabilizing variant of this method using
a hybrid configuration of \afat+\athin to emulate an unreliable processor that
operates close to NTV. From this study, we found that one can improve ETS  even
when the errors slow down the convergence of CG up to 340\%.

As cornerstone of CG method is the matrix-vector product, we believe that
the significance of this study carries over to many other numerical methods for scientific
and engineering applications. On the other hand,
the study has certain limitations. For example,
we did not consider factors such as the
cache hierarchy, interconnection networks, memory buses and bandwidth, which
can be significant in large-scale designs and affect both performance and power consumption. 
We made this choice in order to be able to extract some first-order conclusions 
about the potential of employing NTVC, and 
we intend to investigate those matters in more depth in the future.